\documentstyle[12pt]{article}
\newcommand\beq{\begin{equation}}
\newcommand\eeq{\end{equation}}
\newcommand\bea{\begin{eqnarray}}
\newcommand\eea{\end{eqnarray}}

\begin{document}
\vspace{-2.0cm}
\bigskip

\centerline{\Large \bf Harmonic Oscillator Prepotentials in SU(2)}  
\centerline{\Large \bf  Lattice Gauge Theory} 
\vskip .8 true cm

\begin{center} 
{\bf Manu Mathur}\footnote{Permanent address: S. N. Bose National Centre 
for Basic Sciences, JD Block, Sector III, Salt Lake City, \phantom{iiiiii} Calcutta 
700091, India. E. Mail: manu@bose.res.in}  

Institute of Mathematical Sciences,  Chennai 600 113, India 
\end{center} 
\bigskip

\centerline{\bf Abstract}

We write the SU(2) lattice gauge theory Hamiltonian in 
(d+1) dimensions in terms of prepotentials which are 
the SU(2) fundamental doublets of harmonic oscillators. 
The Hamiltonian in terms of prepotentials has $SU(2) 
\otimes U(1)$ local gauge invariance. In the strong coupling 
limit, the color confinement in this formulation is due to the 
U(1) gauge group. We further solve the $SU(2) \otimes U(1)$ 
Gauss law to characterize the physical Hilbert space in terms 
of a set of gauge invariant integers. We also  obtain 
certain novel gauge invariant operators in terms of the above 
oscillators.  The corresponding  prepotential formulation of SU(N) 
lattice gauge theory is also simple and discussed. 

\vskip .4 true cm

\section{\bf Introduction}

Gauge theories form the underlying framework for both strong and electroweak 
interactions. Therefore, it is desirable and important to understand them  
in terms of their most fundamental structures. In the simplest case of 
SU(2) gauge group, the fundamental operators are the SU(2) doublets. 
All higher representations are built out of them. 
Motivated by this simple fact, we reformulate the SU(2) lattice gauge 
theory in terms of harmonic oscillators (prepotentials) which, under 
gauge transformations, transform as SU(2) doublets. 
\noindent The above construction is based completely on the symmetry 
arguments.  
The SU(2) gauge invariance demands  that 
the physical states satisfy the Gauss law constraint. On lattice, the Gauss 
law involves only electric fields (not vector potentials) and is linear. The 
non-linearity in the Gauss law in the corresponding continuum theory 
reflects itself in the SU(2) commutation relations of the 
lattice electric fields. On the other hand, the SU(2) Lie  algebras and their 
representations are best analyzed using harmonic oscillators through the 
Jordan-Schwinger map \cite{schwinger}. Infact, this harmonic oscillator 
correspondence has been used  in quantum physics and optics 
(in context of SU(2) coherent states) \cite{bellissard}, nuclear physics 
(to describe high spin nucleus states) \cite{marshalek} 
and condensed matter physics (in context of spin chains with global 
SU(2) symmetry) \cite{auer}. However, the Jordan-Schwinger mapping has not 
been exploited in the context of theories with local SU(2) gauge 
invariance. In this work, 
we use this idea to study SU(2) lattice gauge theory by writing the SU(2) 
electric fields in terms Jordan-Schwinger bosons (in the present context, 
we call them prepotentials, see section 3). We find that the 
resulting formulation has the following novel features: 
\begin{enumerate}
\item The Hamiltonian, in terms of prepotentials, has  $SU(2) \otimes U(1)$ 
local gauge invariance. Further, the prepotential doublets also 
form representations of the the new U(1) gauge group. 
\item As expected, the  prepotential SU(2) doublets enable us to study the 
gauge invariance of the theory more critically. In particular, (a) we write 
certain novel gauge invariant operators in terms of prepotentials, (b) in 
this formulation, the physical Hilbert space can be characterized by a set 
of gauge invariant integers. 
\item In the strong coupling limit, it is the U(1) gauge group 
in (1) which is responsible for color confinement.   
\item The matter and prepotentials have similar gauge transformation 
properties and thus are on the same footing. 
\end{enumerate} 
This transformation  to prepotentials  
is also a non-abelian duality transformation and corresponds to transition 
from configuration space of SU(2) rigid rotator to the SU(2) angular 
momentum/electric field basis with the dual Hamiltonian having 
$SU(2) \otimes U(1)$ local gauge invariance. It is known that such duality 
transformations are extremely useful in theories having compact global or local 
symmetries as they extract out the topological degrees of freedom in terms of 
certain integer valued fields and thus isolate 
the effect of compactness. Some simple examples are 2-d xy model \cite{koster}
with global U(1) invariance and pure compact U(1) gauge theories on lattice 
\cite{polyakov}.  
In the latter case, the duality transformations lead to the dual description with 
magnetic monopoles representing the effect of compactness. 
In context of non-abelian lattice 
gauge theories, duality transformations are of special significance 
as it is widely believed that the topological degrees of 
freedom (e.g magnetic monopoles), due to compact nature of the gauge group, 
are responsible for confinement of color \cite{polyakov,thooft1,thooft2}. Towards this 
end, a duality transformation in the context of pure SU(2) lattice gauge in d=2+1 was 
constructed in \cite{sharatramesh,sharat} and was the starting motivation for the 
present work. 
 
\noindent The plan of the paper is as follows: In section (2), we start with an 
introduction to  SU(2) lattice gauge theory in Hamiltonian formulation. This 
section is for the sake of completeness and setting up the notations. 
In Section (3), we describe our prepotential operators and the associated abelian 
gauge invariance. We must emphasize that this abelian gauge invariance is {\it not 
a subgroup} of SU(2). In Section (4), we study the SU(2) gauge transformation 
properties of the prepotentials. The Section (5) is devoted to the study of 
physical Hilbert space in terms of prepotential operators. At the end, we 
briefly mention some of the correponding results for SU(N) lattice gauge 
theories. 

\section{\bf The Hamiltonian Formulation}

\noindent We start with SU(2) lattice gauge theory in (d+1) 
dimensions. The Hamiltonian is:  
\bea 
H = \sum_{n,i} tr E(n,i)^{2} 
+ K \sum_{plaquettes} tr \Big(U_{plaquette} + U^{\dagger}_{plaquette}\Big).    
\label{ham}   
\eea 
where, 
\bea 
U_{plaquette} = U(n,i)U(n+i,j)U^{\dagger}(n+j,i)U^{\dagger}(n,j); 
~~ E(ni) \equiv  E^{a}(n,i){\sigma^{a} \over 2}. 
\nonumber 
\eea 
and K is the coupling constant. The index n labels the site of a d-dimensional 
spatial lattice and i,j (=1,2,...d) denote the direction of the links. Each link (n,i) 
is associated with a symmetric top, whose configuration (i.e the rotation matrix 
from space fixed to body fixed frame) is given by the operator valued SU(2) 
matrix U(n,i) and $E^{a}(n,i) (a=1,2,3)$ are the SU(2) electric field operators 
and can be thought of as the angular momentum operators of the symmetric top in 
the body fixed frame. From now onwards, we associate the angular momentum operator 
$E^{a}(ni)$ to the left of the link (ni). The traces in (\ref{ham}) are over the 
spin half indices of the SU(2) gauge group.  The quantization rules are given by 
\cite{kogut,sharat}: 
\bea 
\left[E^{a}(n,i), U(n,i)\right] = {\sigma^{a} \over 2} U(n,i) => 
\left[E^{a}(n,i),E^{b}(n,i)\right] = - i \epsilon^{abc} E^{c}(n,i) 
\label{cr1} 
\eea    
The Hamiltonian (\ref{ham}) and (\ref{cr1}) are invariant under: 
\bea
 E(ni) \rightarrow V(n) E(ni) V^{\dagger}(n); ~~~ U(ni) \rightarrow V(n)U(ni)V^{\dagger}(n+i). 
\label{gt1} 
\eea 
Thus the generator of left gauge transformation on U(ni) are the electric fields E(ni). The right 
gauge transformations are generated by 
\bea 
e(n,i) \equiv U^{\dagger}(n,i)E(n,i)U(n,i)
\label{sff} 
\eea 
and $e^{a}(ni) (\equiv 2 tr(e(ni)\sigma^{a}/2))$ satisfies: 
\bea 
\left[e^{a}(n,i), U(n,i)\right] = U(ni) {\sigma^{a} \over 2} ; ~~
\left[e^{a}(n,i),e^{b}(n,i)\right] = i \epsilon^{abc} e^{c}(n,i). 
\label{cr2}
\eea 
The fields $e^{a}(ni)$ can be interpreted as the angular momentum operators 
in the space fixed frame of the symmetrical top. They  satisfy the constraints: 
\bea 
\sum_{a=1}^{3} e^{a}(ni)e^{a}(ni) = \sum_{a=1}^{3} E^{a}(ni)E^{a}(ni) 
\label{cons} 
\eea 
at each link (ni).  Under SU(2) transformations: 
\bea 
e(ni) \rightarrow V(n+i) e(ni) V^{\dagger}(n+i)  
\label{gt2} 
\eea 
Therefore, $e^{a}(ni)$ should be associated with the right end of the link (ni).  
The SU(2) Gauss law at every site (n) is: 
\bea 
\sum_{i=1}^{d} \big(E^{a}(n,i) - e^{a}(n-i,i)\big) = 0
\label{gl} 
\eea
The Gauss law (\ref{gl}) states that the sum of all the 2 d angular momenta 
meeting at a site (n) is zero. 

\section{The Prepotentials and Abelian Gauge Invariance} 

We define the SU(2) prepotentials to be two independent doublets of 
harmonic oscillators: 
$(a_{\alpha},a^{\dagger}_{\alpha})$ and 
$(b_{\alpha},b^{\dagger}_{\alpha})$ on  each link (ni). They satisfy: 
\bea 
[a_{\alpha}, a^{\dagger}_{\beta}] = \delta_{\alpha,\beta};~~ [b_{\alpha}, b^{\dagger}_{\beta}] 
= \delta_{\alpha,\beta} ~~~ \alpha,\beta =1,2. 
\label{ho} 
\eea 
Using the Jordan-Schwinger boson representation of SU(2) Lie algebra \cite{schwinger}, 
we write: 
\bea 
E^{a}(n,i) \equiv a^{\dagger}(ni){\tilde\sigma^{a} \over 2} a(ni); ~~~
e^{a}(ni) \equiv b^{\dagger}(ni){{\sigma}^{a} \over 2} b(ni) 
\label{sb} 
\eea 
Note that in (\ref{sb}), ${\tilde\sigma}^{a}_{\alpha\beta} \equiv 
{\sigma}^{a}_{\beta\alpha}$ is used to get the negative sign on the r.h.s of 
the equation (\ref{cr1}).  

\noindent From now onwards, instead of the electric field and the SU(2) link operators 
defined in (\ref{ham}) and (\ref{cr1}), we will treat the harmonic oscillator 
prepotentials in (\ref{ho}) as the basic dynamical variables.  
We generally define the Hilbert space ${\cal H}$ over the configurations of the 
SU(2) rigid rotator. It is obtained by link operators in the various representations 
of SU(2) group, e.g; the physical  Hilbert space is described in terms of gauge 
invariant Wilson loop operators. In the present formulation, the prepotential harmonic 
oscillators create the dual Hilbert space $\tilde{\cal {H}}$ as they are related 
to the SU(2) electric fields. In this sense, what follows should also 
be interpreted as dual formulation of the SU(2) lattice gauge theory.   

\noindent The defining equations (\ref{sb}) for the SU(2) prepotentials imply 
$U(1) \otimes U(1)$  gauge invariance on every link: 
\bea 
a^{\dagger}_{\alpha}(ni) \rightarrow expi\theta(ni)~ a^{\dagger}_{\alpha}(ni); ~~  
b^{\dagger}_{\alpha}(ni) \rightarrow expi\phi(ni) ~b^{\dagger}_{\alpha}(ni).   
\label{u1}
\eea 
In (\ref{u1}), $\theta(ni)$ and $\phi(ni)$ are the arbitrary phase angles at each 
link (ni). Note that the this $U(1) \otimes U(1)$ group is {\it not a subgroup} 
of the SU(2) gauge group. The constraint (\ref{cons}) implies that the 
occupation numbers 
of the harmonic oscillator prepotentials are equal on each link (ni); i.e: 
\bea
\sum_{\alpha=1}^{2} a^{\dagger}_{\alpha}(ni) a_{\alpha}(ni) 
= \sum_{\alpha=1}^{2} b^{\dagger}_{\alpha}(ni) b_{\alpha}(ni)  \equiv N(ni) 
\label{consho} 
\eea 
Therefore,  $\tilde{\cal H}$ of pure SU(2) lattice gauge 
theory is  characterized by the following orthonormal state vectors 
at each link: 
\bea 
| {}^{n~~N-n}_{\bar{n}~~N-\bar{n}} \rangle \equiv {(a^{\dagger}_1)^n  (a^{\dagger}_2)^{N-n} 
(b^{\dagger}_1)^{\bar{n}} (b^{\dagger}_2)^{N-\bar{n}} \over {\sqrt{n!}  
\sqrt{(N-n)!} \sqrt{\bar{n}!} \sqrt{(N-\bar{n})!}}} | {}^{0~~0}_{0~~0} \rangle . 
\label{hs}
\eea 

\noindent The invariance of $\tilde{\cal H}$ under (\ref{u1}) implies that the 
gauge group $U(1) \otimes U(1)$ reduces to U(1) with $\theta(ni) 
= - \phi(ni)$. The constraint  (\ref{consho}) now becomes the Gauss law 
for this resulting abelian gauge invariance.

\section{The SU(2)  Gauge Invariance} 

Under the SU(2) gauge transformations (\ref{gt1} and \ref{gt2}): 
\bea 
a^{\dagger}_{\alpha}(ni) \rightarrow V(n)_{\alpha\beta} a^{\dagger}_{\beta}(ni), ~~~
b^{\dagger}_{\alpha}(ni) \rightarrow b^{\dagger}_{\beta}(ni)(V^{\dagger}(n+i))_{\beta\alpha}  
\label{gt3} 
\eea 
Thus, the two sets of prepotentials  transform like SU(2) doublets, one 
from the left and the other from the right. We further define: 
$\tilde a_{\alpha} \equiv \epsilon_{\alpha,\beta} a_{\beta}$ 
and $\tilde b_{\alpha} \equiv \epsilon_{\alpha,\beta} b_{\beta}$. 
Under SU(2) gauge transformations,  $\tilde a_{\alpha}$ and $\tilde b_{\alpha}$ 
transform as $a^{\dagger}_{\alpha}(ni)$ and $b^{\dagger}_{\alpha}(ni)$ respectively. 
Exploiting the above symmetry properties, we now directly write down the operator 
valued SU(2) matrix U(ni) in the Hilbert space ${\tilde{\cal H}}$ as:  
\bea
U(ni)_{\alpha\beta} = F(ni)  \big(a^{\dagger}(ni)_{\alpha} b^{\dagger}(ni)_{\beta} 
+ \tilde{a}(ni)_{\alpha} \tilde{b}(ni)_{\beta}\big) F(ni)  
\label{dh}
\eea
In (\ref{dh}), $F(ni) \equiv  {1 \over \sqrt{N(ni)+1}}$ with N(ni) defined in (\ref{consho}). 
It is the normalization factor and is required for the operator valued SU(2) 
matrix to be unitary. More explicitly on a particular link: 
\bea 
{U} ~= {1 \over \sqrt{a^{\dagger}.a + 1}} ~ \left( \begin{array}{cc} 
a^{\dagger}_{1}b^{\dagger}_{1} + a_2 b_2 &  a^{\dagger}_{1}
b^{\dagger}_{2} - a_2 b_1   \\
a^{\dagger}_{2}b^{\dagger}_{1} - a_1 b_2 & a^{\dagger}_{2}b^{\dagger}_{2} + a_1 b_1  
\end{array}\right)~ {1 \over \sqrt{a^{\dagger}.a + 1}} \nonumber 
\eea
This is the standard structure of a SU(2) matrix; i.e it is of the form: 
$ U = \left( \begin{array}{cc} z_1 &  -z_2^{\dagger}   \\
z_2 & ~~z_1^{\dagger} \end{array}\right)$ with $z_1^{\dagger}z_1 + z_2^{\dagger}z_2 = 
z_1 z^{\dagger}_1 + z_2 z^{\dagger}_2 =1$ on $\tilde{\cal H}$.  Further, it can be 
explicitly checked that the operator valued matrix elements of U in (\ref{dh}) commute 
amongst themselves and (\ref{dh}) is consistent with the defining equation (\ref{sff}) for the 
generator of the right gauge transformations $e(ni)$.  
The dual form of the SU(2) link operators in (\ref{dh}) can also 
be derived by the use of the Wigner Eckart theorem on the $\tilde{\cal H}$. The 
first and second terms in (\ref{dh}) change the value of the angular momentum by 
$+ {1 \over 2}$ and $- {1 \over 2}$ units respectively\footnote{In terms of the 
Young tableau, the first term corresponds to adding a new 
box in the horizontal row and the second term corresponds to deleting a box from the 
horizontal row.}. Note that (\ref{dh}) is invariant under under the new U(1) 
gauge transformation 
(\ref{u1}) with $\theta(ni) = - \phi(ni)$.  Thus we have broken the SU(2) link operators 
U(ni) into the left $(a_{\alpha}(ni))$ and the right $(b_{\alpha}(ni))$ transforming 
prepotentials. This separation will help us to construct the gauge invariant states 
of the theory in the next section. 
The Hamiltonian in (\ref{ham}) can now be written in it's dual form: 
\bea
H =  \sum_{ni} {N(ni) \over 2} \Big({N(ni) \over 2} + 1 \Big) 
+ \sum_{plaquettes} tr \Big(U_{plaquette} + U^{\dagger}_{plaquette}\Big).    
\label{dham}
\eea 
The first term in (\ref{dham}) depends on the number operator  on all 
the links of the lattice. The second term is made up of the 4 links of the plaquettes 
given by (\ref{dh}). The dual Hamiltonian in (\ref{dham}) is trivially invariant 
under the $SU(2) \otimes U(1)$ gauge transformations. In this dual formulation,  
we can write new gauge invariant operators which one can not write in terms of 
the original link fields. To see this we define 
two operators:  
\bea
U^{+{1\over 2}}(ni)  \equiv a^{\dagger}(ni)_{\alpha} b^{\dagger}(ni)_{\beta}, ~~~ 
U^{-{1\over 2}}(ni)  \equiv \tilde{a}(ni)_{\alpha} \tilde{b}(ni)_{\beta}  
\label{gio} 
\eea 
and notice that 1) $ U^{\pm{1\over 2}}(ni)$ have  the same gauge transformation 
properties as $U(ni)$, 2) they are both invariant under the U(1) gauge transformation. 
Infact, these are the two terms appearing in (\ref{dh}).  
Therefore, one can define the fundamental gauge invariant operators as consisting of 
products of $ U^{\pm{1\over 2}}(ni)$ over the links of a directed closed loop.  The 
standard Wilson loops in this dual language can be written as the sum of these basic 
gauge invariant operators but the reverse is not possible. 
Further, the coupling of matter with the gauge fields 
is also simple in this formulation. Let $(q^{\dagger}_{\alpha}(n),q_{\alpha}(n))$ 
be the doublets of  matter creation annihilation operators at site (n). Under the SU(2) 
gauge transformations they transform as: $q_{\alpha}(n) \rightarrow V_{\alpha\beta}(n)q_{\beta}(n)$  
The $SU(2)\otimes U(1)$ singlet interaction terms are: $(q^{\dagger}.a^{\dagger})(b^{\dagger}.q)$ 
and $(q^{\dagger}.\tilde{a})(\tilde{b}.q)$. Note that these  matter-prepotential  couplings 
are novel as the minimal coupling with the original link variable $U_{\alpha\beta}$ is 
only the sum of the two. Therefore, it is inadequate to incorporate all the possible gauge 
invariant interactions. 

\section{The Physical Hilbert Space ${\tilde {\cal H}}^{p}$} 

In this section, we exploit the simple gauge transformation properties of the 
prepotentials to characterize the $SU(2) \otimes U(1)$ invariant physical 
Hilbert space denoted by ${\tilde{\cal H}}^{p}$. We first solve the SU(2) Gauss 
law. For this purpose, it is convenient to 
collect the set of 2d prepotential creation operators  associated with the site (n) 
as: $c^{\dagger}(n,\bar{i})$ with $\bar{i} =1,2,....,2d$,  where, 
\bea 
c^{\dagger}(n,i)  & \equiv & \tilde{a}^{\dagger}(n,i) \nonumber \\
c^{\dagger}(n,d+i)  & \equiv &  {b}^{\dagger}(n-i,i); ~~ i=1,2,...,d. \nonumber     
\eea 
\noindent With the above relabelling, one can easily check:     
\begin{enumerate} 
\item The new prepotentials also satisfy harmonic oscillator algebra: 
$\left[c_{\alpha}(n,\bar{i}),c^{\dagger}_{\beta}(n,\bar{j})\right] = 
\delta_{\alpha,\beta} \delta_{\bar{i},\bar{j}}$. 
\item The SU(2) Gauss law (\ref{gl}) at site n is: 
\bea 
J^{a}_{total}(n) \equiv \sum_{\bar{i}=1}^{2d} c^{\dagger}(ni)_{\alpha} \left({\sigma^{a}
\over 2}\right)_{\alpha\beta} {c}_{\beta}(ni)    = 0  
\label{dm}
\eea 
and it simply states that the sum of all the angular momenta meeting at the site (n),
${\vec J}_{total}(n)$, is zero. 
\item Under SU(2) gauge transformations, all the 2d operators $c^{\dagger}(n,\bar{i})$ 
transform from the right as SU(2) doublets. 
\end{enumerate}   
Thus, the problem of constructing the most general SU(2) gauge invariant states at 
site (n) reduces to constructing SU(2) singlets out of 2d spin half prepotentials  
$c^{\dagger}_{\alpha}(n\bar{i})$.  We denote the physical Hilbert space 
at site n, consisting of all such invariants by $\tilde{\cal H}_{n}^{p}$. 
Therefore, $\tilde{\cal H}_{n}^{p}$ is characterized as: 
\bea
|\vec{l}(n)> \equiv \left\vert \begin{array}{cccc} 
l_{12} & l_{13}  & ... & l_{1~2d}     \\
 & l_{23} & ... & l_{2~2d} \\
 &  & . & . \\
 &  &  & ~~~~~l_{2d-1~2d} \\
\end{array} \right \rangle =  \prod_{{}^{\bar{i},\bar{j}}_{\bar{j} > 
\bar{i}}} \left(c^{\dagger}(n\bar{i}).
\tilde{c}^{\dagger}(n\bar{j})\right)^{l_{\bar{i}\bar{j}}(n)}
|0> . 
\label{giv} 
\eea 
In (\ref{giv}), $l_{ij}(n) \left(\equiv l_{ji}(n)\right)$ are $N_{d} = d(2d-1)$ 
 +ve integers which are invariant under the SU(2) gauge transformations. 
The states (\ref{giv})  $ \in \tilde{\cal H}_{n}^{p}$
are the eigenvectors of $c^{\dagger}(n\bar{i}).c(n\bar{i})$, $\bar{i}=
1,2,..,2d$, with eigenvalue $\left(\sum_{\bar{j} \neq \bar{i}} 
l_{\bar{i}\bar{j}}(n)\right)$. 
Therefore, the operator $\prod_{{}^{\bar{i},\bar{j}}_{\bar{j} >
\bar{i}}} \left(c^{\dagger}(n\bar{i}).
\tilde{c}^{\dagger}(n\bar{j})\right)^{l_{\bar{i}\bar{j}}}$ in (\ref{giv}) 
creates ${1 \over 2} \left(\sum_{\bar{j} \neq \bar{i}} 
l_{\bar{i}\bar{j}}(n)\right)$ units of electric flux on the link 
$(n\bar{i})$. Thus, in terms of the prepotentials, 
the problem of solving the non-abelian SU(2) Gauss law reduces to the 
problem of solving U(1) Gauss law (see also \cite{sharatgna}). 
The complete $SU(2) \otimes U(1)$ invariant Hilbert space can be written as: 
\bea 
{\tilde{\cal H}}^{p}  = \prod_{n}{}^{\prime} \otimes {\tilde{\cal H}}_{n}^{p}
\label{phs} 
\eea 
In (\ref{phs}), the direct product is taken over all the lattice sites such that 
U(1) Gauss law (\ref{consho}) is satisfied. 
Note that this construction is dual 
description of the construction of the gauge invariant states through the Wilson 
loop operators acting on the vacuum.  Further, like Wilson loops, (\ref{giv}) describes 
an over complete basis. Therefore, the set of  $N_{d}$ integers is not the minimal 
set required to characterize ${\tilde{\cal H}}_{n}^{p}$. Infact, the minimum 
set consists of a set of $N_{d}-N_{d-1}-d = 3(d-1)$ integers per lattice site. 
These gauge invariant integers are associated with the physical transverse 
degrees of freedom of the three SU(2) gluons.  

\noindent To see the color confinement in this formulation, we imagine  
a quark q(n) and anti-quark $\bar{q}(n)$ pair located at lattice sites n and m 
respectively. Following the previous sections, we can construct the  
SU(2) color 
invariant states {\it locally} at (n) and (m) as  $(b^{\dagger}.q)$ and 
$(q^{\dagger}.a^{\dagger})$ respectively. It is the U(1) gauge invariance 
which forces us to connect them through a string of harmonic oscillator 
prepotentials and get a $SU(2) \otimes U(1)$ gauge invariant physical state. 
In the strong coupling limit, the energy of this configuration is proportional 
to the length of the string which is forced by the U(1) gauge invariance. 
This is probably related to the idea of 
t' Hooft \cite{thooft2} that in SU(N) gauge theory the $U(1)^{N-1}$ group is 
relevant for the color confinement mechanism. Infact,  the prepotential 
formalism in the case of  SU(N) lattice gauge theories should have  2 (N-1) 
sets of prepotentials,  transforming as 
(N-1) fundamental representations of SU(N) \cite{manu2}. Therefore, this 
formalism is invariant under  $SU(N) \otimes \left(U(1)\right)^{N-1}$ gauge 
group. This work will be reported elsewhere \cite{manu2}. 

{\section {Discussion and Summary}} 

\noindent 
The complete set of commuting operators in pure SU(2) lattice gauge theory  
are $E^{a}(ni)E^{a}(ni)$, $E^{3}(ni)$ and $e^{3}(ni)$. 
The common eigenvectors are denoted 
by $|j, m, \bar{m}> $ where j,m and $\bar{m}$ denote the eigenvalues of the above three 
operators respectively. We note that the states in (\ref{hs}) can also be constructed 
using the standard link operators $U_{\alpha\beta}$. Infact, 
\bea 
|j,m,\bar{m}\rangle = \sum_{i_1,i_2,....i_{2j} \in S_{2j}} U_{m_{i_1}\bar{m}_1} 
U_{m_{i_2} \bar{m}_2}...U_{ m_{i_{2j}} \bar{m}_{2j} } |0 \rangle 
\label{st}
\eea 
In (\ref{st}), $(m_1,m_2,..,m_{2j})$ and $(\bar{m}_1,\bar{m}_2,..,\bar{m}_{2j})$ 
are the two sets of  $\pm {1 \over 2}$  with constraints:  
$m_1+m_2 +..m_{2j} = m$, $\bar{m}_1+\bar{m}_2 +..\bar{m}_{2j} = 
\bar{m}$ and $S_{2j}$ is the permutation group of order 2j. The correspondence with 
(\ref{hs}) is: $2j = N, m = 2n-{N\over 2}$ and  $\bar{m} = 2\bar{n}-{N\over 2}$.
The construction (\ref{st}) becomes more and more complicated  as j increases. 
Thus, the the characterization of the Hilbert space of the lattice gauge theories 
through the prepotential formulation presented in this work is much simpler than 
the standard formulation. 

In this work, we have presented a new platform in terms of harmonic 
oscillator prepotentials to analyze the non-abelian lattice gauge theories. 
The formulation is in terms of the dynamical variables which  belong to the 
most fundamental representation(s) of the gauge group. The novel features are 
already summarized in section (1). 

\vspace{0.6cm} 

\noindent {\bf Acknowledgements} 

\vspace{0.4cm} 

\noindent It is a pleasure for me  to thank H.S. Sharatchandra for the invitation 
to visit Institute of Mathematical Sciences (IMSc), Chennai. I also want to 
thank Ramesh Anishetty, N. D. Haridass and H.S Sharatchandra for various discussions. 
In particular, the discussions with Ramesh Anishetty and H. S. Sharatchandra in the 
context of \cite{sharatramesh} and their comments and suggestions on the initial 
manuscript were useful. Thanks to  Naveen, Raghu, Santosh and Vinu for the pleasant 
time at IMSc. Also, IMSc hospitality is happily acknowledged.

\end{document}